# Robust Sound Source Tracking Using SRP-PHAT and 3D Convolutional Neural Networks

David Diaz-Guerra, *Student Member, IEEE,* Antonio Miguel and Jose R. Beltran

***Abstract*—In this paper, we present a new single sound source DOA estimation and tracking system based on the well-known SRP-PHAT algorithm and a three-dimensional Convolutional Neural Network. It uses SRP-PHAT power maps as input features of a fully convolutional causal architecture that uses 3D convolutional layers to accurately perform the tracking of a sound source even in highly reverberant scenarios where most of the state of the art techniques fail. Unlike previous methods, since we do not use bidirectional recurrent layers and all our convolutional layers are causal in the time dimension, our system is feasible for real-time applications and it provides a new DOA estimation for each new SRP-PHAT map. To train the model, we introduce a new procedure to simulate random trajectories as they are needed during the training, equivalent to an infinite-size dataset with high flexibility to modify its acoustical conditions such as the reverberation time. We use both acoustical simulations on a large range of reverberation times and the actual recordings of the LOCATA dataset to prove the robustness of our system and its good performance even using low-resolution SRP-PHAT maps.**

***Index Terms*—microphone arrays, direction of arrival estimation, DOA, sound source tracking, SRP-PHAT, convolutional neural networks, CNN.**

## I. Introduction

DIRECTION Of Arrival (DOA) estimation and Sound Source Localization with microphone arrays has been widely investigated and used in different applications, such as robot audition [1], [2], acoustic characterization [3], speech recognition [4], [5] or teleconference systems [6]. Most of the techniques in the literature can be roughly classified into i) Time Difference Of Arrival (TDOA) based techniques, which first use the Generalized Cross-Correlation (GCCs) functions [7] to estimate the TDOA and then compute the most reliable DOA for them (it is worth saying that there are also some alternatives to the GCCs such as the eigenvalue decomposition [8] or even deep-learning based techniques [9]), ii) beamforming based techniques, such as SRP-PHAT [10], [11], which search the direction that maximizes the power of the output of a beamformer, and iii) subspace techniques, such as Multiple Signal Classification (MUSIC) [12], [13], based on the eigenstructure of the narrowband cross-correlation matrices. These techniques vary in terms of computational complexity and their robustness against adverse scenarios such as noise and reverberation. When they have to deal with non-stationary signals, such as the speech, a tracking algorithm is needed after them to exploit the temporal correlation between source positions [14]–[17].

D. Diaz-Guerra, A. Miguel and J.R. Beltran are with the Department of Electronic Engineering and Communications, University of Zaragoza, Zaragoza, Spain.

More recently, some methods based on Deep Neural Networks (DNN) have been proposed. From the original Multilayer Perceptron (MLP) used in the first proposals [18], [19], their architectures have evolved into more sophisticated Convolutional Neural Networks [20]–[22] which can jointly perform DOA estimation and tracking. However, despite DNN-based techniques claim to be more robust than the classical methods, their use of the CNNs might not be the most appropriate and many of them add non-causal recurrent layers that make them unfeasible for real-time applications.

In this paper, we propose the use of 3D CNNs over SRP-PHAT power maps to jointly perform the DOA estimation and the tracking of a source in highly reverberant rooms. We present a completely causal technique that provides a new DOA estimation with each new power map and we show its robustness through several simulations in adverse conditions. We analyze how the resolution of the SRP-PHAT power maps affects our technique and we prove that by using CNNs we can obtain resolutions that surpass the search grid employed to compute the maps. Finally, we apply our model to the acoustic source LOCalization And TrAcking (LOCATA) challenge [23] dataset in order to show how the models trained with simulated signals are general enough to work with actual recordings in real conditions. Although we focus on compact arrays and evaluate the performance of our technique with an array with 12 microphones mounted over a NAO robot head, the technique may be used with any array geometry.

It is worth mentioning that we focus on single source scenarios that are supposed to always have an active source; therefore, our tracking does not need to deal with data association and with the birth and the dead of the source. For a single source scenario, the birth and dead problem may be easily solved with a Voice Activity Detector (VAD) but extending our system to deal with multiple sources might be more difficult. However, some ideas are proposed in section III-A.

In order to encourage and facilitate the replicability of this research, the source code of our model and the models used as baselines, as well as everything needed to train and test them, can be found in our public repository[1]; we also share the trained models there.

The remainder of this paper is structured as follows. We first review the SRP-PHAT algorithm (section II) and the state of the art of DNN-based DOA estimation techniques (section III). In section IV we present our proposed technique and in section V we analyze its performance with both simulated rooms and actual recordings. Finally, section VI concludes the paper.

[1] https://github.com/DavidDiazGuerra/Cross3D

## II. THE SRP-PHAT ALGORITHM

The signal received at the $n^{th}$ sensor of a microphone array can be modeled as

$$x_n(t) = a_s(t) * h_n(\boldsymbol{\theta}_s, t) + v_n(t), \tag{1}$$

where $a_s(t)$ is the signal generated by the source, $\boldsymbol{\theta}_s$ is the source position, $h_n(\boldsymbol{\theta}_s, t)$ is the impulse response from $\boldsymbol{\theta}_s$ to the $n^{th}$ sensor, and $v_n(t)$ is the noise of the sensor, which is typically supposed to be white, Gaussian, and uncorrelated with the source signal and with the noises of other sensors. It is worth mentioning that $\boldsymbol{\theta}_s$ is written in bold because it can represent an angle, two spherical coordinates, or even a point in 3D Cartesian coordinates depending on the geometry of the array.

One of the most classic and popular approaches to DOA estimation is finding the direction that maximizes the Steered Response Power (SRP) that we would obtain using a filter-and-sum beamformer:

$$\hat{\boldsymbol{\theta}}_s = \underset{\boldsymbol{\theta}}{\operatorname{argmax}}\, P(\boldsymbol{\theta}) \tag{2}$$

$$P(\boldsymbol{\theta}) = \int_{-\infty}^{+\infty} \left| \sum_{n=0}^{N-1} G_n(\omega) X_n(\omega) e^{-j\omega\tau_n(\boldsymbol{\theta})} \right|^2 d\omega, \tag{3}$$

where $N$ is the number of sensors of the array, $X_n(\omega)$ is the Fourier Transform of $x_n(t)$, $G_n(\omega)$ is the frequency response of the filter for the channel $n$, and $\tau_n(\boldsymbol{\theta})$ is the time delay occurring from the position or direction $\boldsymbol{\theta}$ to the $n^{th}$ sensor.

Although directly implementing (3) would be computationally expensive, it can be computed in terms of the Generalized Cross-Correlation functions as

$$P(\boldsymbol{\theta}) = 2\pi \sum_{n=0}^{N-1} \sum_{m=0}^{N-1} R_{nm}(\Delta\tau_{nm}(\boldsymbol{\theta})), \tag{4}$$

where $\Delta\tau_{nm}(\boldsymbol{\theta}) = \tau_n(\boldsymbol{\theta}) - \tau_m(\boldsymbol{\theta})$ and $R_{nm}$ is the GCC between the signals of the $n^{th}$ and the $m^{th}$ sensor:

$$R_{nm}(\tau) = \frac{1}{2\pi} \int_{-\infty}^{+\infty} \Psi_{nm}(\omega) X_n(\omega) X_m^*(\omega) e^{j\omega t} d\omega, \tag{5}$$

where $^*$ denotes the complex conjugate and $\Psi_{nm}(\omega) = G_n(\omega) G_m^*(\omega)$ is a weighting function.

Equation (4), combined with the use of the PHAse Transform (PHAT) $G_n(\omega) = 1/|X_n(\omega)|$, is commonly known as the SRP-PHAT algorithm [10], [11], and allows us to obtain an acoustic power map of the environment whose maximum should correspond with the source position.

Although the SRP-PHAT algorithm is a good trade-off between robustness and computational efficiency, obtaining more accurate results than two-step TDOA based techniques with a lower computational cost than most of the broadband subspace techniques, it still presents several issues. The main advantage of (4) is that most of its computational cost is in computing the GCCs and does not increase with the search space. However, the computation of its sums for each direction, especially if it is needed to interpolate $R_{nm}(\Delta\tau_{nm}(\boldsymbol{\theta}))$ from its adjacent samples, may not be negligible; this problem becomes more challenging when the search space is two-dimensional, e.g. the two angles of the spherical coordinates, or even three-dimensional, e.g. XYZ coordinates. Some search strategies have been proposed to reduce the number of evaluations of (4) that need to be computed to accurately find the maximum of $P(\boldsymbol{\theta})$ [24]–[26] but, due to the non-convexity of the SRP-PHAT power maps, the number of SRP-PHAT evaluations needed might still be an issue in some scenarios. In [26], [27], it is proposed to modify (4) to compute the power received from a space region instead of from a point, so they can use hierarchical search strategies over maps with lower resolution.

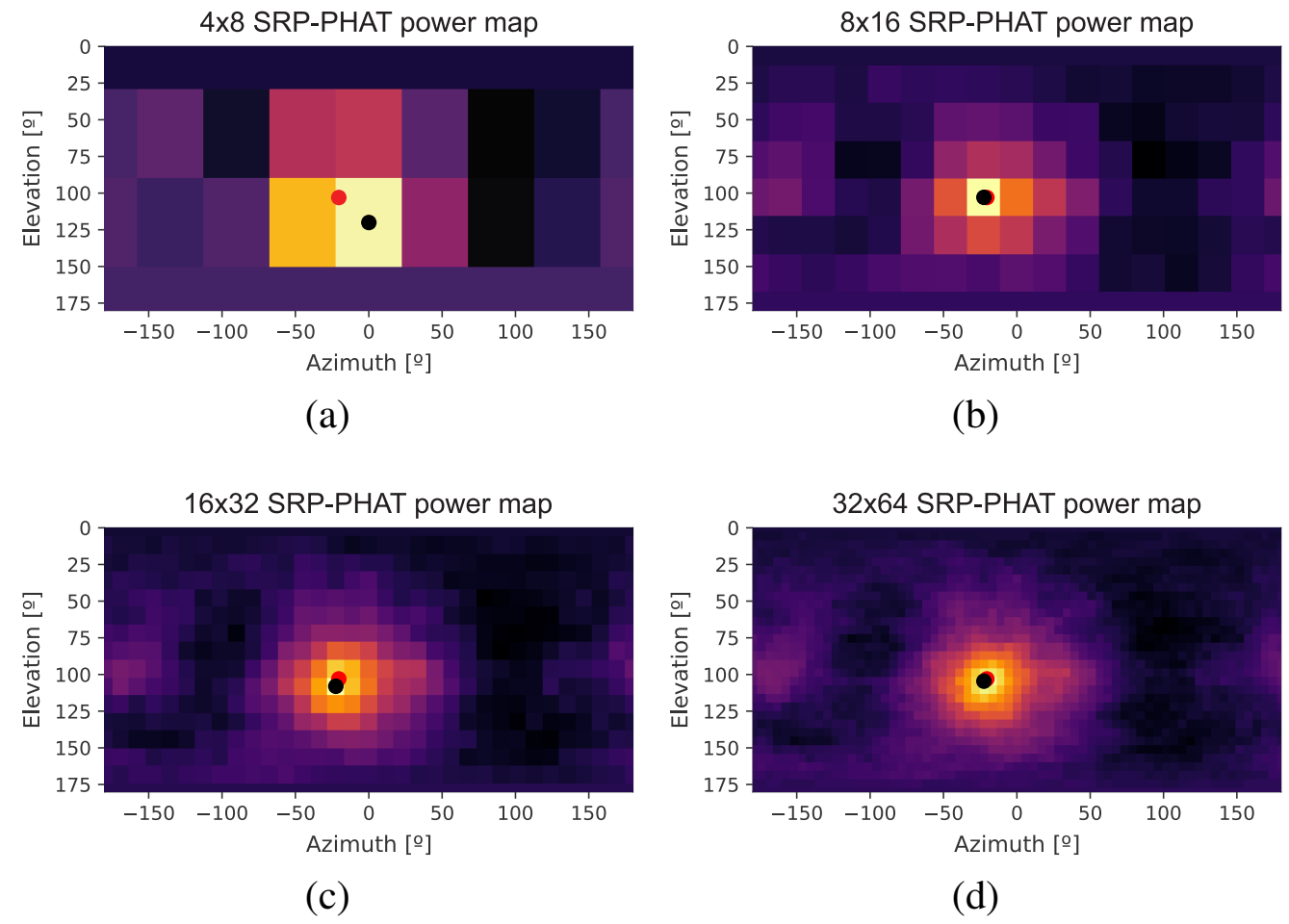

Fig. 1. Example of SRP-PHAT power maps with different resolutions in a favorable scenario: SNR=30 dB and $T_{60}$=0.3 s. The red dot indicates the actual DOA of the sound source and the black dot is at the maximum of the map.

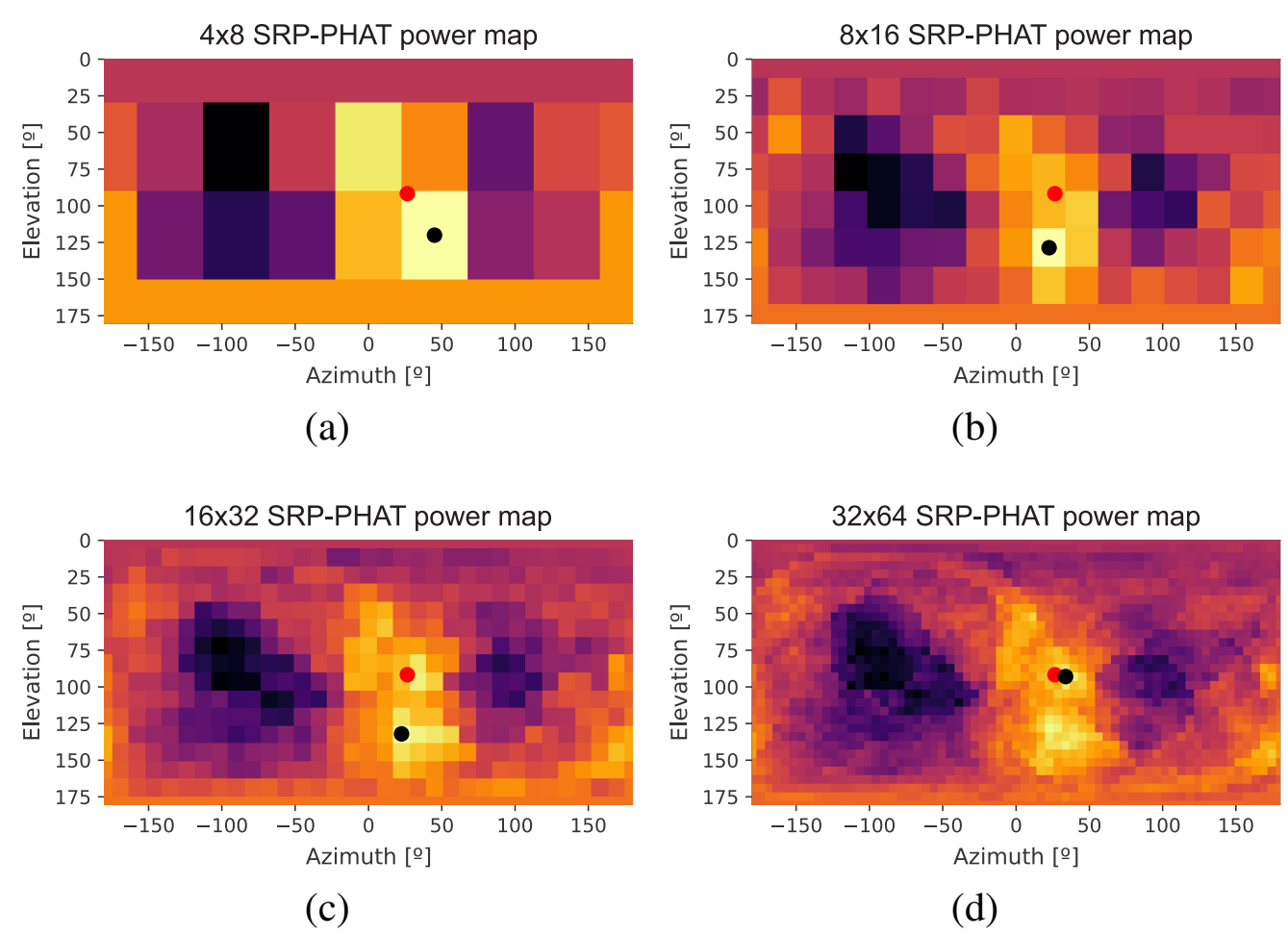

Fig. 2. Example of SRP-PHAT power maps with different resolutions in an adverse scenario: SNR=5 dB and $T_{60}$=0.9 s. The red dot indicates the actual DOA of the sound source and the black dot is at the maximum of the map.

As we can see in Fig. 1, in favorable scenarios with high SNR and low reverberation, the SRP-PHAT power maps have a clear maximum in the DOA of the sound source that can be used to obtain a good estimation even with low-resolution maps but, when SNR decreases and the reverberation increases, such as in the scenario of Fig. 2, the maps present several local maxima that may be incorrectly interpreted as the DOA of the sound, especially when using low-resolution maps. However, in those maps, in addition to the maxima, we can also observe several patterns that are also related to the DOA of the sound and the geometry of the array and that may be exploited to obtain a more accurate DOA estimation.

Due to the non-stationary nature of most of the signals of interest, such as the speech or the music, a tracking stage is needed after the DOA estimation to exploit the temporal correlation between the source positions and to avoid inaccurate estimations in frames where the power of the signal is low or its autocorrelation makes the maximum of the power map become too wide. The algorithms for one source tracking are typically based on the Kalman filter [14], [15] although more advanced techniques have been proposed to deal with multiple sources, such as those based on particle filtering [16], [17]. However, in these approaches, they use two-step strategies which make them sensitive to potential information loss when only the DOA estimations are used for the tracking; e.g. the absolute maximum of the SRP-PHAT maps is always selected even if another local-maximum was much closer to the previous estimations and we assign the same likelihood to all the DOA estimations while some of them correspond to wider maximums from frames where the source was weaker. Including some of this information in the tracking algorithms may be possible, but it would increase both its complexity and the number of parameters that would need to be fine-tuned. In [28], a technique to share information between an iterative DOA estimator based on Expectation-Maximization [29] and a tracking system based on particle filtering is proposed. In this paper, we use Neural Networks to jointly perform DOA estimation and tracking, since they have been proved to have an excellent performance in several end-to-end problems in other fields, such as computer vision or speech recognition or synthesis.

## III. DOA estimation with Deep Neural Networks

One of the first proposals of using neural networks for DOA estimation was [18], which used a fully connected perceptron with a hidden layer to obtain the DOA from the GCCs as a classification problem. Since that, several techniques have been proposed, differing in the output format, the features that they use as input, and the network architecture.

### A. Output format

To obtain the DOA estimation as a classification problem, we first need to define a grid of directions where the source can be found (similar to the resolution of the SRP-PHAT maps) so the network has an output per grid point. The network of [18] had 359 outputs, so they had a maximum resolution of 1 degree for azimuth estimation. However, if we want to estimate both azimuth and elevation (or even XYZ coordinates), the number of outputs would dramatically increase, and therefore its computational complexity and the size of the dataset needed to train the network.

In a footnote of [18], it was claimed that they had obtained worse results when they tried to estimate the DOA as a regression problem and classification approaches seem to be the most popular [19], [30]–[34]. However, since [22] proposed using 3 outputs to estimate a unitary vector pointing to the direction of the source, several models have followed this approach [35]–[37]; further studies about the advantages of using 3 regression outputs to infer the Cartesian coordinates instead of 2 to directly obtain the azimuth and the elevation can be found in [38], [39]. Motivated by the good results of these recent works, we opted to follow a regression approach.

One of the main drawbacks of solving the DOA estimation as a regression problem is that it makes it harder to estimate the DOA of multiple sources. Since they also classify the sources into several classes, Sound Event Localization and Detection (SELD) models usually have a regression output for each source class [22], [35]–[37] supposing that only one source of each class can be active at the same time. Another possible approach might be the use of a single-source DOA estimator (as the one proposed here) combined with a source cancellation technique [40], [41] to iteratively find multiple sources.

### B. Input features and network architecture

Initially, the most common input features were the GCCs between the signals of each sensor [18], [33], [34], but we can also find other approaches such as using the eigenvectors of the spatial covariance matrix [19]. In [42], we proposed using low-resolution SRP-PHAT maps, in that case combined with fully connected perceptrons. More recently, some techniques have been proposed using 2D convolutional networks over the spectrogram of the microphone signals, using only the phase information [30], the magnitude information [32] or both of them [22], [31], [35]; some transformations, such as using the cepstrogram [21] or the Mel spectrogram [37], have also been proposed. Other features proposed as inputs of CNNs are Ambisonics intensity vectors [43], raw audio samples [20] and combinations of several of the already mentioned features [21], [36], [37].

One of the most important properties of CNNs is that they are equivariant to translations, which, in plain text, means that if we apply a translation to the input features we get the same output with its equivalent translation. This property is very useful in many computer vision applications, where the same patterns have the same meaning in all the positions of the image. When used for DOA estimation and tracking, 2D CNNs are typically used over spectrograms, so convolution is performed over the time and the frequency axes and each microphone spectrogram is treated as a different channel. Being equivariant to time translations seems to be an advantage since we would expect similar patterns for any source in a position no matter the time instant when it was there. However, since the phase differences for the same source position vary with the frequency, equivariance to frequency shifts may not be an interesting property. Another approach to the use of 2D CNNs is proposed in [30], where the convolution is performed over the frequency and the microphone dimensions and the time evolution is not taken into account by the network, i.e. they do not perform any tracking. As they work with an Uniform Linear Array (ULA), the phase differences expected for a source position are the same for each pair of adjacent microphones, so equivariance is desired, but this would not be the case for other array geometries.

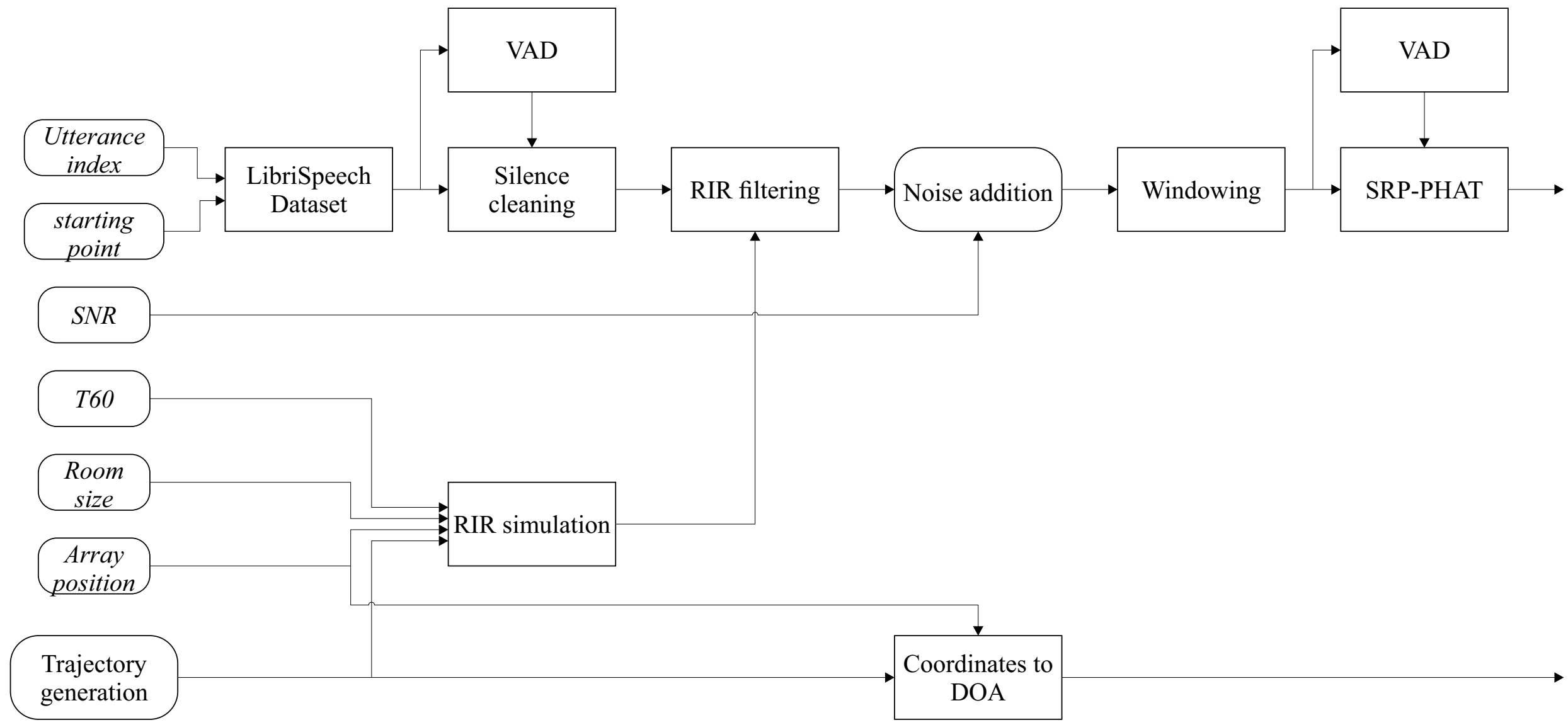

Fig. 3. Dataset generation process. Italic letters represent variables and regular letters represent processes. Right-angled boxes represent deterministic processes and round boxes represent stochastic variables or processes.

In this paper, we propose the use of CNNs over SRP-PHAT power maps, performing the convolution over the dimensions of the maps and the temporal dimension. Any kind of SRP-PHAT power maps could be employed with this approach depending on the geometry of the array, but as we focus on compact arrays, we use 2D spherical power maps and therefore, since we include the temporal dimension, 3D CNNs. Actually, working over spherical maps, equivariance to spherical translations (i.e. rotations) would be preferred over equivariance to euclidean translations, but this would lead us to the use Spherical CNNs [44], which are still less efficient from a computational point of view than classical CNNs. The extension to 4D CNNs over 3D SRP-PHAT maps to perform 3D Sound Source Localization (SSL) with distributed arrays would be straightforward.

Many of the state of the art CNN architectures include bidirectional recurrent units at the last layers of the model. Recurrent Neural Networks (RNNs), as recurrent linear filters, make the output at any time instant dependent on the values of the input at every previous time instant and, therefore, applying them in the backward direction is extremely non-causal. Obviously, any tracking system can benefit greatly from the information of the future positions of the source but, in order to make our system feasible for real-time applications, we opt for using only causal convolutional layers.

## IV. Proposed technique

### A. Training dataset

Due to the difficulty of obtaining an accurately hand-labeled dataset of moving sources recorded with microphone arrays, we opted to train our model with simulated signals; another approach might have been using measured RIRs convolved with speech signals, but this would have reduced the amount of different acoustic conditions seen by the model during training, increasing the possibility of overfitting to those conditions and not generalizing. Instead of generating a dataset and using it to train the network, we simulate the inputs of the networks as they are needed during training. This makes the training slower, but has two important advantages: 1) we have an infinite-size dataset, since all the random parameters of the simulation are modified for each trajectory simulated during the training, which reduces the risk of overfitting, and 2) we have higher flexibility to modify the probability distribution of the parameters of the simulation, such as the signal to noise ratio or the reverberation time, during training so we can perform curriculum learning strategies [45].

As shown in Fig. 3, we use LibriSpeech utterances as sound sources. The LibriSpeech corpus [46] contains 960 hours of speech sampled at $f_s = 16\,\mathrm{kHz}$ extracted from audiobooks. Although audiobooks could be expected to contain quite clean speech signals, we found that some of them have a strong background noise that, after filtered by the RIRs, would be located in the same position as the source and would facilitate its localization and tracking in silent segments. To avoid our network to learn to exploit this fact, which will not be present in actual recordings, we use the WebRTC Voice Activity Detector (VAD) [47] to detect silent segments and clean them by completely removing the signal in those frames.

The size of the rooms are randomly selected from the range $3\,\mathrm{m} \times 3\,\mathrm{m} \times 2.5\,\mathrm{m}$ to $10\,\mathrm{m} \times 8\,\mathrm{m} \times 6\,\mathrm{m}$ and the array is randomly placed inside the room, being restricted to have a separation from the walls of a 10% of the room size in each dimension and be in the lower half of the room for the vertical axis. The Signal to Noise Ratio (*SNR*) and reverberation time ($T_{60}$) are also randomly selected from the ranges $5\,\mathrm{dB}$ to $30\,\mathrm{dB}$ and $0.2\,\mathrm{s}$ to $1.3\,\mathrm{s}$ respectively. Uniform distributions over the specified ranges are used for all the random parameters of the dataset.

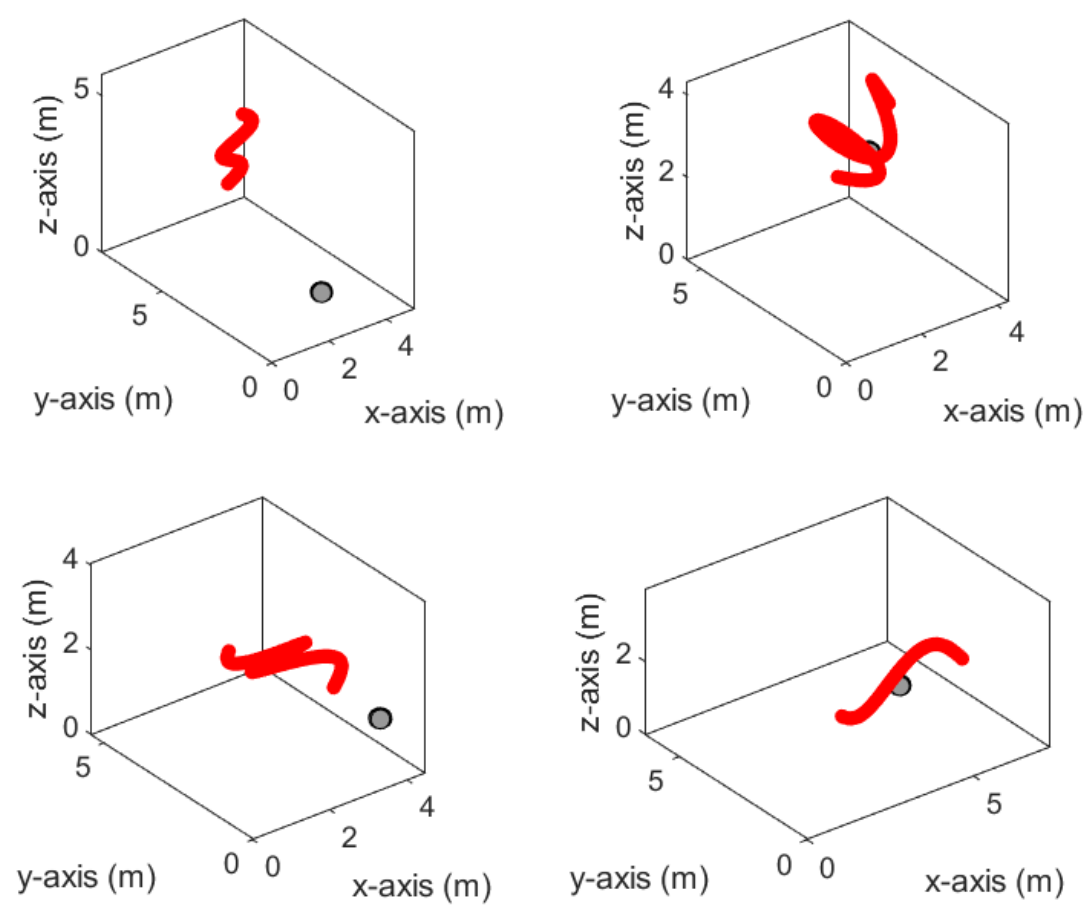

Fig. 4. Examples of source trajectories used to train the model. The red dots are the trajectory points and the gray points represent the microphones.

We need to randomly generate continuous trajectory points, so it is possible to track them, but having enough diversity to avoid the network to learn how they are generated and overfit to them. In order to do so, we randomly select two points within the room boundaries to be the starting ($\mathbf{p}_0 = [p_{x0}, p_{y0}, p_{z0}]^T$) and ending ($\mathbf{p}'_L = [p'_{xL}, p'_{yL}, p'_{zL}]^T$) points of the trajectory and add to the straight line that connect them a sinusoidal function in each axis with random frequencies ($\boldsymbol{\omega} = [\omega_x, \omega_y, \omega_z]^T$) and amplitudes ($\mathbf{A} = [A_x, A_y, A_z]^T$) ensuring that no more than 2 oscillations are performed during the trajectory in each axis and that the amplitude is low enough to avoid the source to exit the room:

$$\mathbf{p}_i = \mathbf{p}_0 + \frac{i}{L-1}(\mathbf{p}'_L - \mathbf{p}_0) + \mathbf{A} \circ \sin(\boldsymbol{\omega} i), \qquad (6)$$

where $L$ is the number of points of the trajectory, $\circ$ stands for the pointwise product and the $\sin$ function also operates pointwise. Although the generation model is quite simple, it generates quite diverse trajectories (some examples are shown in Fig. 4) and, since the network only sees the azimuth and elevation coordinates and has a limited temporal perceptive field, the model should not overfit to it. In order to confirm this, we tested our model in a more realistic scenario with the recordings of the LOCATA dataset (see section V-B2).

To simulate the movement of the source, we use the GPU implementation of the Image Source Method [48] found in the python library gpuRIR [49]; the use of this library allows us to reduce the simulation time in two orders of magnitude and makes possible to perform the simulations during the training of the network.

For the results presented in this paper, we simulated a microphone array with 12 sensors designed to be mounted over a NAO robot head; the minimum and maximum inter-microphone distance of the array are $1.3\,\mathrm{cm}$ and $12.1\,\mathrm{cm}$ and the actual position of each microphone can be found in the documentation of the LOCATA dataset [50]. We did not include the effect of the robot head in the simulation. Including the scattering generated by it might improve the results obtained with the LOCATA dataset, but would have also increased the complexity of the simulations and would made the training slower. Since different array geometries would lead to different patterns, the model should be re-trained for any new microphone array. We could take advantage of the similarity of the power maps of most compact arrays to apply transfer learning strategies when training models for new arrays. However, this is not a big issue since we do not need to record a dataset with the new array but just simulate it.

Having simulated the propagation of the sound from the moving source to each microphone of the array using the overlap-add method, we add an omnidirectional Gaussian noise to obtain the desired SNR, window the signal using Hanning windows of length $K = 4096$ samples (i.e. $256\,\mathrm{ms}$) with a hop size of $3K/4$, and apply (4) to obtain the SRP-PHAT map of each window. In order to compute the noise power needed to obtain the desired SNR, we computed the signal power as the average power of all the non-silent frames of the trajectory. Finally, we subtract its mean to each map and divide it between its maximum to fit it to the range [-1,1]. For the sake of computational efficiency, we do not perform any kind of interpolation in the computation of (4) and just approximate the fractional delays to the nearest sample.

We found that, since this simulation process did not include any directional noise, the models trained with it were very sensitive to directional noise sources. For example, in some of the recordings of the LOCATA dataset, the noise of a fan is present and, although its power is very low, the models tracked it when it was the only active sound source. In order to avoid this issue, we use the WebRTC VAD to determine in which frames the speech source is active. We first tried to include the VAD information as an additional input channel to the network. However, as during the training the VAD sometimes failed and classified frames which speech information as silent, the network learned that even the frames classified as silent could contain useful tracking information as long as they contained a directional source and therefore ignores the VAD input. In order to avoid the network to track the directional noise sources of the LOCATA dataset, we finally opted to turn to zero the maps corresponding to frames classified as silent by the VAD so no directional information was seen by the network when there was not any speech source active.

### B. Model architecture

Our model takes as input a 4-dimensional tensor ($\mathbf{M}$) with size $C \times T \times N_\theta \times N_\varphi$, whose first channel $\mathbf{M}_{1,t,i,j}$ is built by computing $T$ temporally equispaced SRP-PHAT maps with $N_\theta$ equispaced elevation angles in the range $\theta \in [0, \pi]$ and $N_\varphi$ equispaced azimuth angles in $\varphi \in [-\pi, \pi)$; for planar arrays, the same model could be used sampling the elevation only in $\theta \in [0, \frac{\pi}{2}]$. Using uniform spherical sampling instead of equispaced angles might have led to more precise SRP-PHAT maps, but would prevent us from using standard convolutional layers.

Although the model must learn more complex patterns in order to exploit all the information available in the SRP-PHAT maps, it is obvious that one of the main sources of information about the DOA of the source is the position of the maximum of each map; however, the $\operatorname{argmax}$ function is highly non-linear (and non-differentiable) and it is not easy for an artificial neural network to learn and fit it. Since it did not cause a significant increase of the computational complexity of the algorithm, we decided to explicitly indicate to the network the position of the maximum of each map. After trying to introduce this information in different layers, we found that the best results were obtained including it in the input of the network, using $C = 3$ with $\mathbf{M}_{2,t,i,j} = \hat{\theta}_t^{SRP}$ and $\mathbf{M}_{3,t,i,j} = \hat{\varphi}_t^{SRP}$ for any $t \in \{1, ..., T\}$, $i \in \{1, ..., N_\theta\}$, and $j \in \{1, ..., N_\varphi\}$, where $\hat{\theta}_t^{SRP}$ and $\hat{\varphi}_t^{SRP}$ are the DOA equivalent to the position of the maximum of the map $t$ normalized to be in the range [0,1]. This approach might seem quite redundant and inefficient, but it is a typical approach to condition the output of a CNN to the value of a variable since it is the simplest way to include that information in the first layers of the network keeping its convolutional architecture, whose implementation is extremely optimized in the Deep Learning software libraries.

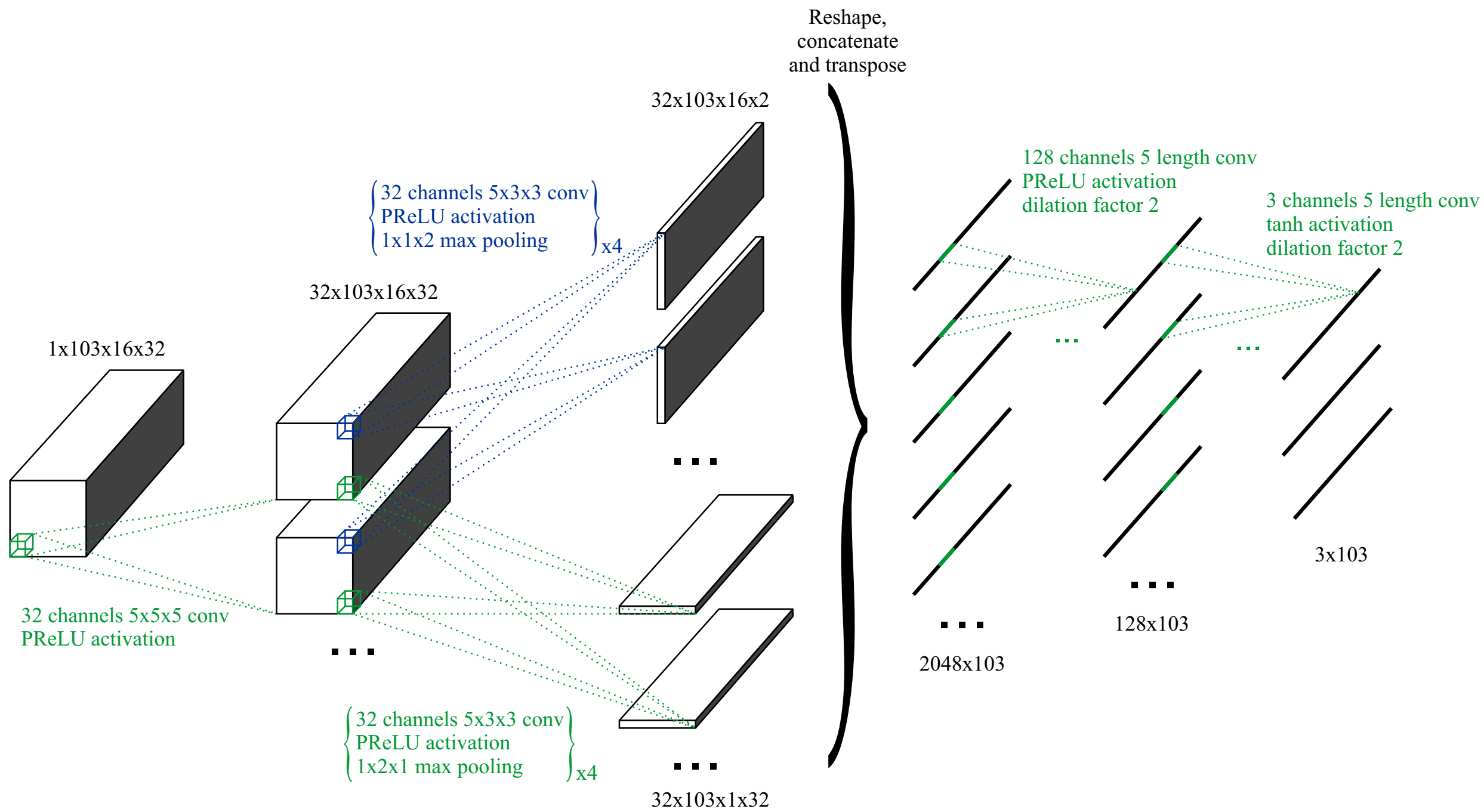

Fig. 5. Model architecture. The noted sizes correspond to a model for 16x32 maps and an input sequence of length 103. For the shake of simplicity, we represented it with only 1 input channel, although it actually have 3.

The first layer of our model is a 3D convolutional layer with 32 kernels of size $5 \times 5 \times 5$ and PReLU activations [51]. It is worth mentioning that, for the temporal axis, we always use causal convolutions, so this model could be used in real time applications generating a new DOA estimation for each new power map available and without introducing any delay.

Pooling layers are typically used in CNNs to progressively reduce the size of the input and make the model generalize; not using them, means that the fully connected layers used at the end of most of the convolutional models would have a huge number of trainable parameters which would surely overfit. When the desired output of the CNN is a summary of the information of the whole input, e.g. in image classification tasks, increasing the number of channels with convolutional layers and reducing their size with pooling layers progressively reduce the spatial information and gets higher-level representations of the input. However, since our desired output is not only related to the presence of some patterns but especially to their position, we must be careful when using them.

In order to get the benefits of pooling layers but allowing the spatial information to reach the last layers of the model, we opted to, as shown in Fig. 5, split the model into two branches and apply max pooling in a different dimension in each one. Working this way, the branch which pools the $\varphi$ axis can retain positional information about the $\theta$ coordinate of the maps and vice versa. Specifically, each branch has 4 layers with a convolution with 32 kernels of size $5 \times 3 \times 3$, PReLU activations, and a max-pooling with a kernel size of $1 \times 1 \times 2$ and $1 \times 2 \times 1$ respectively. If the input power maps have less than 16 points in the $\theta$ or the $\varphi$ axes, it would not be possible to perform so many pooling layers; in those cases, we reduce the 4 layers to the maximum number possible: $\log_2 (\min (N_\theta, N_\phi))$. Due to the use of 3D convolutional layers and these perpendicular branches, we named our model Cross3D.

After the 3D convolutional layers, we concatenate the results of each branch and reshape them so we have a temporal sequence of length $T$ for each one of the elements of each channel and spherical coordinates. Each one of these temporal sequences are used as the input channels of a 1D causal convolutional layer with 128 kernels of length 5 and PReLU activations. Finally, the resultant 128 time sequences are passed through another 1D causal convolutional layer with only 3 kernels of length 5 and $\tanh$ activations. These layers are similar to the fully connected layers that most of the CNN architectures have, but we include a temporal convolution so they can still exploit the tracking information. We use a dilation factor of 2 in order to allow the tracking to take into account a longer context without increasing the complexity of the network. With all the temporal convolutions included in the model, each DOA estimation is computed from the last 37 SRP-PHAT maps, i.e. the tracking memory is $7.17\,\mathrm{s}$.

TABLE I
MODELS EMPLOYED FOR THE EVALUATION

| Model | Input | | Trainable parameters | Temporal perceptive field | Window length | Causal |
|---|---|---|---|---|---|---|
| Cross3D | Power maps | 4x8 | 526 372 | 5.63 s | 4096 | Yes |
| | | 8x16 | 946 340 | 6.40 s | | |
| | | 16x32 | 1 693 988 | | | |
| | | 32x64 | 5 626 148 | 7.17 s | | |
| | | 64x128 | 21 354 788 | | | |
| 1D CNN | GCCs | | 11 282 436 | 7.17 s | 4096 | Yes |
| | Maximums (64x128) | | 6 899 716 | | | |
| 2D CNN | Spectrograms | | 1 882 372 | 7.17 s | 4096 | Yes |
| SELDnet [22] | | | 104 643 | $\infty$ | 512 | No |

The result of all this process is 3 time sequences of length T whose elements are in the range $(-1, 1)$, which are considered to be the XYZ coordinates of a unitary vector pointing in the direction of the source in each time frame. Tables detailing the network architecture of Cross3D for several SRP-PHAT map resolutions can be found in the supplementary material of this paper.

### C. Training

We trained our model to minimize the Euclidean distance between the output of the network and the 3 time sequences obtained from the coordinates of the unitary vectors that point to the direction where the sound source was simulated in each time window. Similarly to the results reported in [22], [38], [39], we obtained better results using this approach than trying to directly obtain the spherical coordinates from the network even when using the great-circle distance between the output and the ground-truth DOA angles as cost function.

Although using an infinite-size dataset the term "epoch" does not have the same meaning than in most of the machine learning systems, we define an epoch as 585 trajectories (the number of book chapters in the LibriSpeech train-clean-100 subset). We employed 80 epochs with trajectories of $20\,\mathrm{s}$, i.e. 103 SRP-PHAT maps, to train the model with the Adam algorithm [52] using Pytorch [53].

As explained in section IV-A, we trained the networks with reverberation times and SNRs uniformly distributed from $0.2\,\mathrm{s}$ to $1.3\,\mathrm{s}$ and from $5\,\mathrm{dB}$ to $30\,\mathrm{dB}$ respectively; however, we found that the training converged faster with higher SNRs. Therefore, we followed a curriculum learning strategy [45] using batches of 5 trajectories with SNR=$30\,\mathrm{dB}$ for the first 20 epochs and for the following epochs we employed the full range of SNRs, increased the batch size to 10 trajectories, and reduced the learning rate from 1e-4 to 1e-5.

## V. EVALUATION

### A. Baseline methods

In order to analyze the convenience of using SRP-PHAT maps as input features of CNNs for DOA estimation, we developed some alternative CNNs to use them as baseline. We designed them to be as similar as possible to our proposed model and to have the same temporal perceptive field so they have the same tracking information.

Since we are including the position of the maximum of each map into the input of the network, we should verify if our model is actually exploiting the additional information that is within the SRP-PHAT maps or if it is only using the position of its maximums. To do that, we designed a 1D CNN which takes as input 2 time sequences with the coordinates of the maximum of each map normalized to the range [0,1] and applies to them 7 layers of 1D causal convolutions with PReLU activations and without any pooling. All the layers had a kernel size of 5 and the last two layers used a dilation factor of 2, so its temporal perceptive field is 37 frames as in Cross3D, and the number of channels of each layer was $\{1024, 512, 512, 512, 512, 128, 3\}$. The results shown in the following sections were obtained training this network with the same process described in section IV-C and using the coordinates of maps with resolution 64x128.

One of the most common input features employed by the first DOA estimation techniques based on neural networks were the GCCs. They typically employed fully connected perceptrons with not too many hidden layers and, since they only used the GCCs computed in a temporal window, did not perform any kind of tracking. Following this idea, but with the aim of including tracking information to the network, we used the same 1D causal CNN than we used over the map maximums but using as input sequences the temporal evolution of each element of the GCCs which represented an inter-microphone delay lower than the maximum inter-microphone distance divided by the speed of sound.

Although, as explained in section III, the use of 2D CNNs over spectrograms may not be optimal, we also implemented a model following this approach since it is quite popular in the literature. For a fair comparison, we used causal convolutions with a similar architecture to Cross3D: one convolution with 256 5x5 kernels, four convolutions with 256 5x5 kernels with 1x4 pooling, a reshape to transform the remaining features into temporal sequences and two 1D causal convolutions with kernel size 5 and dilation factor 2, the first one with 128 channels and the last one with 3. For computing the spectrogram, we used the same windows than for computing the SRP-PHAT maps, extracted the magnitude and phase of each frequency of the FFT, and finally normalized the magnitude of each window to its maximum and the phase to the range [-1,1].

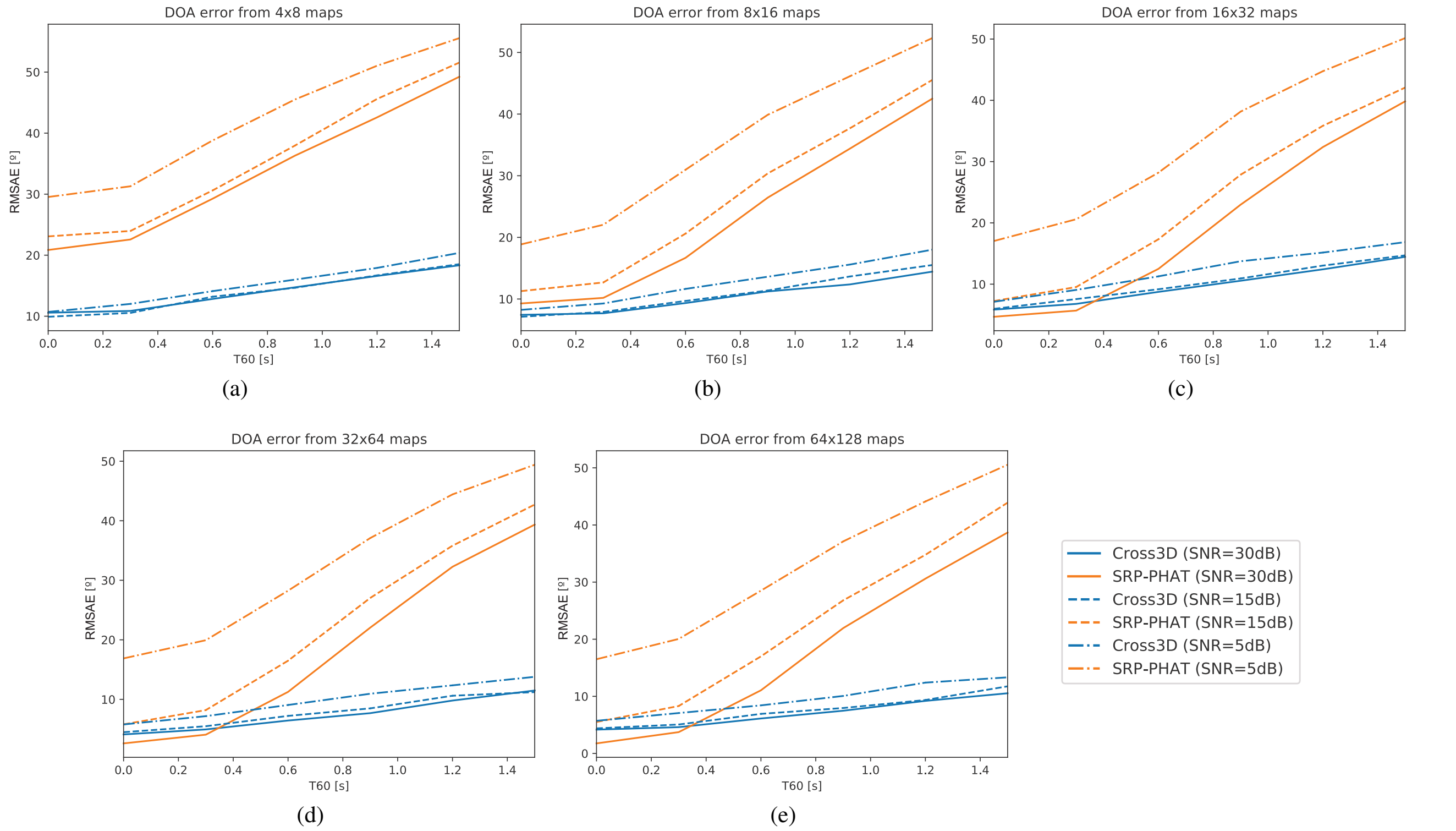

Fig. 6. Localization Root Mean Squared Angular Error for several power map resolutions, SNRs and reverberation times. The silent frames where not included in the computation of the RMSAE.

Finally, we also trained with our simulation procedure a replica of SELDnet [22] but without including the Sound Event Detection (SED) output and with only a DOA output since we were only interested in tracking one source. This model takes as inputs the magnitude and phase of the spectrograms and has three 2D convolutional layers followed by two bidirectional Gated Recurrent Units (GRU) [54] and two fully connected layers. It is worth saying that this model, due to the bidirectional GRUs, is non-causal and that it uses shorter analysis windows than the other analyzed methods.

For the models that use spectrograms as input features, we found that they did not train properly with the full range of reverberations described in section IV-A, and we got the best results training them with values of T60 randomly selected from the range $0\,\mathrm{s}$ to $0.3\,\mathrm{s}$.

All the models employed for the evaluation are summarized in Table I and tables detailing their architectures can be found in the supplementary material of this paper.

## B. Results

*1) Simulated dataset:* We trained different models for several power map resolutions with the whole range of reverberation times and SNRs, and then we tested their performance for several specific values of $T_{60}$ and SNR in order to analyze the robustness of the proposed tracking system.

Since we are using SRP-PHAT power maps as the input of our algorithm, we started our evaluation comparing our model with the classic SRP-PHAT algorithm. SRP-PHAT does not perform any kind of tracking, so, for a fairer comparison, we did not take into account the silent frames when computing the Root Mean Squared Angular Errors (RMSAE) showed in Fig. 6. As we can see in this figure, when working with high-resolution power maps in almost anechoic rooms with high SNR, using our 3D CNN over the SRP-PHAT maps does not improve the results compared to just taking the maximum of each map; actually, our system seems to slightly degrade the DOA estimation, probably due to the effect applying an unneeded tracking. However, when the room conditions deteriorate, we can see how Cross3D is robust enough to get its performance degraded in only $5^\circ$ when the $T_{60}$ increases to $1.5\,\mathrm{s}$ (which is higher than any reverberation seen during the training) while the SRP-PHAT algorithm is just unable to perform a proper estimation. They are also surprising the results obtained with maps of only 4x8 resolution, which only perform a SRP-PHAT measurement each $45^\circ$ in the azimuth and $60^\circ$ in the elevation and, since $P(\theta, \varphi) = P(\theta, 0)$ $\forall \varphi \in [0, 2\pi)$ if $\theta = 0$ or $\pi$, only needs to perform 18 computations of (4).

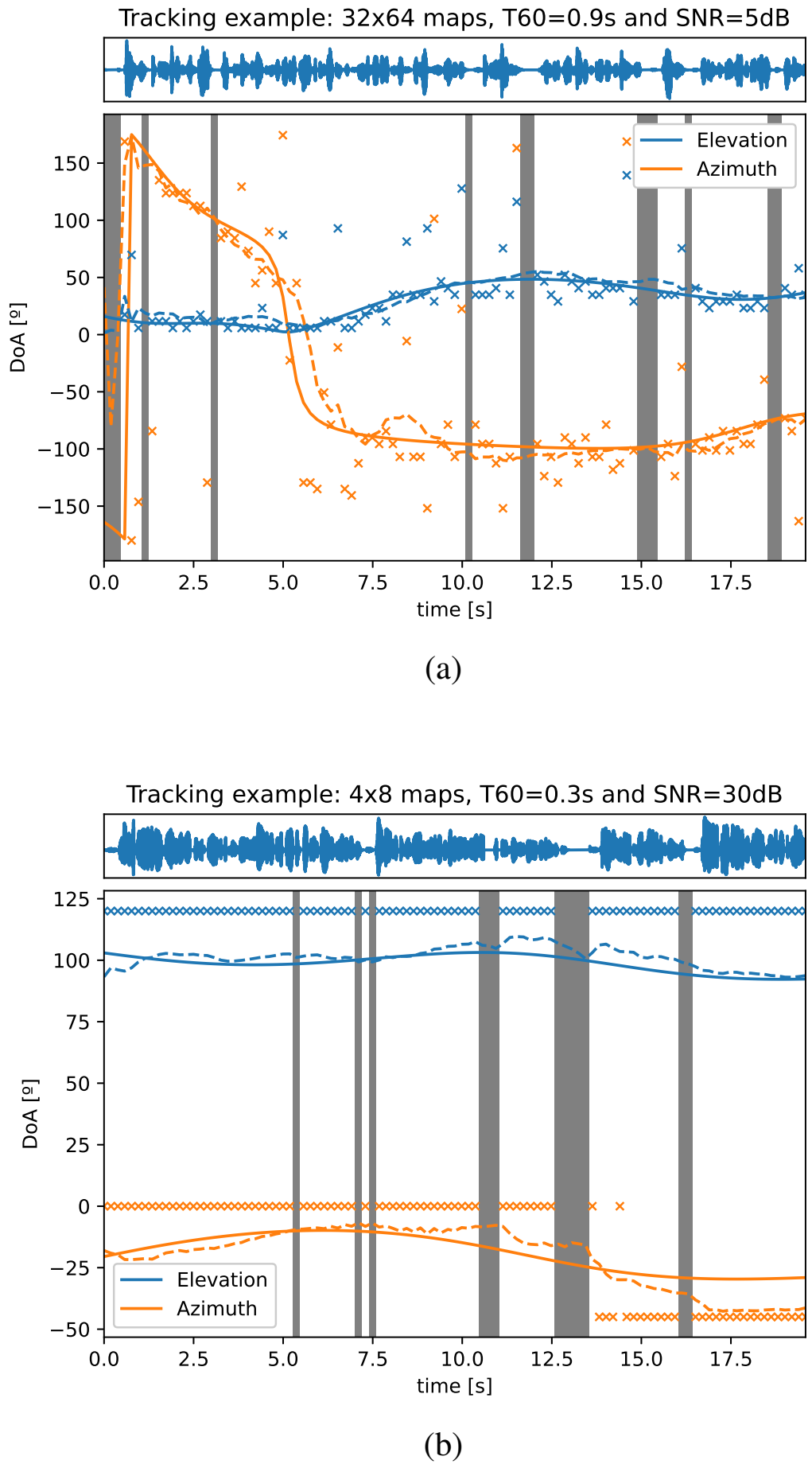

Fig. 7. Examples of the DOA estimated in a scenario with $T_{60} = 0.9\,\mathrm{s}$ and SNR=5 dB using maps with 32x64 resolution (a) and in a scenario with $T_{60} = 0.3\,\mathrm{s}$ and SNR=30 dB using maps with 4x8 resolution (b). The solid line represents the actual DOA of the source, the dashed line the estimated DOA and the crosses represent the maximum of each SRP-PHAT power map. Grey segments indicate silent frames.

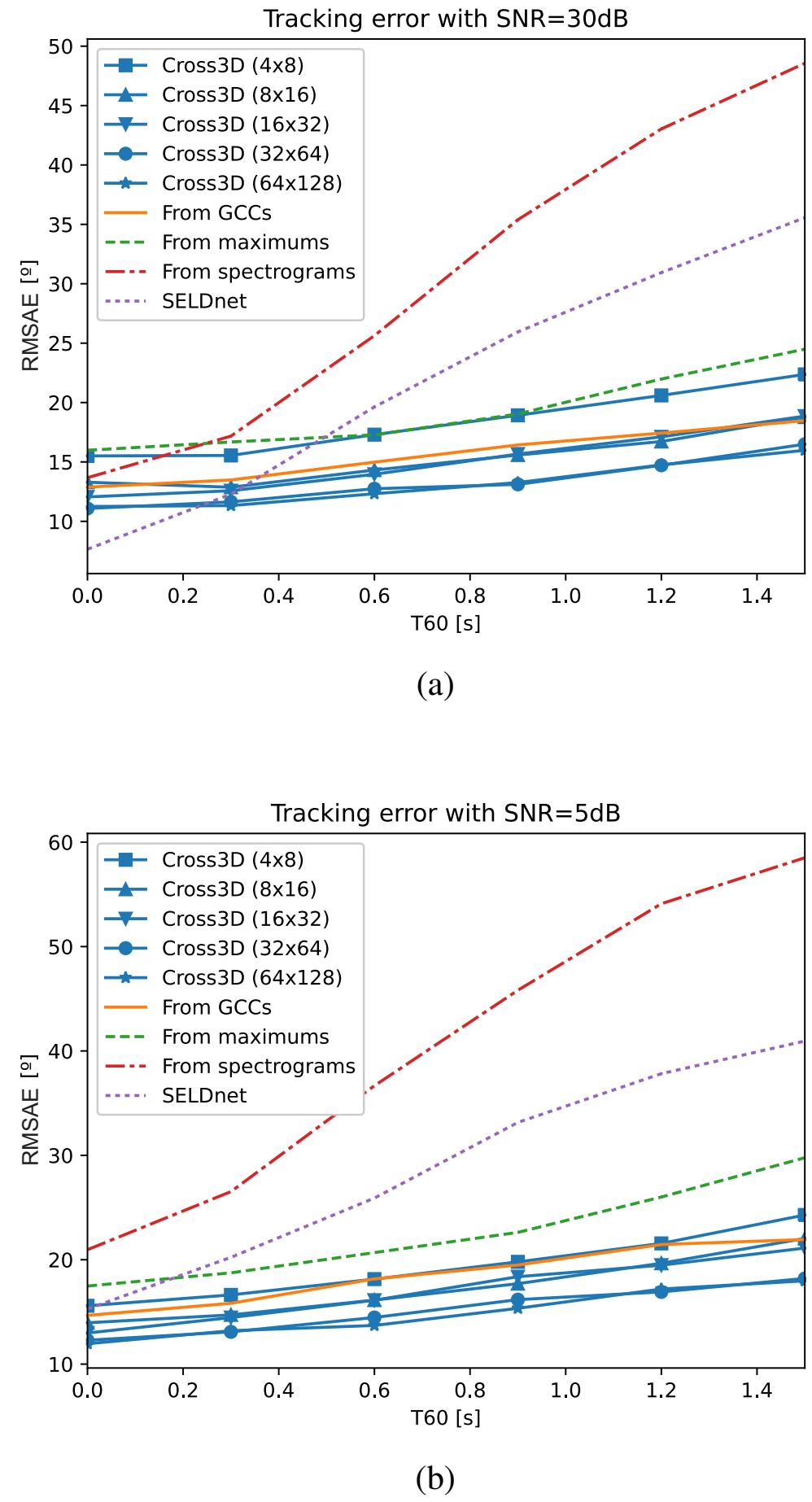

Fig. 8. Tracking Root Mean Squared Angular Error of Cross3D with several power map resolutions and the baseline methods for SNR=30 dB (a) and SNR=5 dB (b) and several reverberation times. The silent frames where also included in the computation of the RMSAE.

Fig. 7 shows a couple of examples of simulated trajectories and their estimated DOA. In Fig. 7a we can see how, for scenarios with high reverberation and low SNRs, the maximum of the SRP-PHAT maps becomes really noisy but our proposed system is able to maintain the estimated DOA quite close to the actual one. In linear systems, robust tracking with noisy estimations usually comes with the cost of being slow to track fast changes, at least with casual systems, but we can see how our model was able to follow the sudden change in the azimuth of the source at the fifth second of the trajectory. In Fig. 7b we can see how, when working with low-resolution power maps, our system is able to predict the DOA with much higher precision than the maximums of the maps. This could not be done with a two-step DOA estimation and tracking algorithm that performed the tracking based only on the maximum of the maps. Our system is able to analyze the whole maps and it was able to learn to exploit the patterns in the SRP-PHAT maps to achieve higher resolution than the grid used to compute the maps.

Finally, we also tested the baseline methods under different reverberations and SNRs to compare the robustness of each model. In this case, since all the methods include tracking capabilities, we did not exclude the silent frames when we computed the RMSAEs shown in Fig. 8. We can see how the best results are obtained using our method with high resolution power maps, but that, even reducing the resolution, its performance is still competitive. Using 1D CNN over the coordinates of the maximums of 64x128 SRP-PHAT maps performs worse than using our 3D CNN over 4x8 maps, so we can conclude that our model is exploiting the patterns present in the SRP-PHAT maps and not only using the information of the position of its maximums (this was also suggested by Fig. 7b). Using 1D CNNs over the GCCs —which is an approach, to the best of the authors' knowledge, unpublished— have a performance between using 3D CNNs over 4x8 and 8x16 maps and may be an interesting approach when a lower computational cost is needed. Finally, the models that use spectrograms as inputs perform well in favorable scenarios (SELDnet even outperforms our proposal in low noise anechoic chambers) but they are not very robust against noise and reverberation.

TABLE II
RMSAE [°] OF THE DOA ESTIMATED FOR THE LOCATA DATASET WITH CROSS3D USING SEVERAL MAP RESOLUTIONS AND THE BASELINE TRACKING METHODS. THE SILENT FRAMES WERE INCLUDED IN THE COMPUTATION OF THE RMSAE.

| | Model: | Cross3D | | | | | 1D CNN | | 2D CNN | SELDnet |
|---|---|---|---|---|---|---|---|---|---|---|
| | Input: | SRP-PHAT maps 4x8 | 8x16 | 16x32 | 32x64 | 64x128 | GCCs | Maximums | Spectrograms | |
| Task 1 | Recording 1 | 17.93 | 11.92 | 8.30 | 4.62 | 5.16 | 16.18 | 7.54 | 93.76 | 29.70 |
| | Recording 2 | 18.90 | 7.68 | 6.68 | 4.90 | 3.91 | 12.60 | 5.19 | 64.18 | 38.44 |
| | Recording 3 | 10.35 | 6.34 | 2.98 | 3.25 | 2.24 | 11.57 | 5.09 | 140.21 | 54.81 |
| | Average | 15.72 | 8.65 | 5.99 | 4.26 | 3.77 | 13.45 | 5.94 | 99.38 | 40.98 |
| Task 3 | Recording 1 | 23.06 | 18.11 | 13.79 | 12.43 | 9.92 | 13.59 | 14.04 | 70.86 | 50.57 |
| | Recording 2 | 20.97 | 13.71 | 10.01 | 8.36 | 9.22 | 14.17 | 12.02 | 83.42 | 48.71 |
| | Recording 3 | 21.05 | 12.74 | 9.83 | 7.69 | 6.60 | 15.21 | 13.29 | 82.48 | 57.29 |
| | Average | 21.69 | 14.85 | 11.21 | 9.49 | 8.58 | 14.32 | 13.12 | 78.92 | 52.86 |
| Task 5 | Recording 1 | 11.93 | 10.83 | 7.25 | 5.74 | 5.49 | 10.93 | 10.53 | 58.33 | 37.24 |
| | Recording 2 | 20.92 | 16.16 | 16.08 | 12.18 | 13.59 | 17.33 | 17.42 | 41.98 | 73.17 |
| | Recording 3 | 23.57 | 18.25 | 13.58 | 15.64 | 15.49 | 20.14 | 23.58 | 66.91 | 66.50 |
| | Average | 18.81 | 15.08 | 12.31 | 11.19 | 11.52 | 16.13 | 17.18 | 55.74 | 58.97 |
| Average | | 18.74 | 12.86 | 9.83 | 8.31 | 7.96 | 14.64 | 12.08 | 78.01 | 50.94 |

*2) LOCATA dataset:* In order to confirm that, although it was trained with a simulated dataset, our system is general enough to track sound sources recorded in real rooms, we tested it with the LOCATA challenge dataset [23], which contains several recordings with the same array that we had simulated to train the models. We used the development dataset and we focused in the tasks 1, 3, and 5 of the challenge: a static loudspeaker recorded with a static array, a moving talker recorded with a static array, and a moving talker recorded with a moving array; it is worth mentioning that the array was static in all the simulations employed to train the model. For the robot head microphone array that we simulated in the training dataset, the development dataset contains 3 recordings for each task and its ground-truth positions.

It is worth saying that the only modification to the proposed technique that we made after seeing its performance with the LCOATA dataset was the use of a VAD. All the hyperparameters of the model and the acoustical properties of the training dataset were selected according only to the results obtained with simulated datasets. In other words, we used simulated signals for training and validation and the LOCATA recordings only for testing.

Table II shows the RMSAE of estimating the DOA of the source of each recording using our technique and with the baseline methods. Although it is difficult to draw conclusions from such a low number of recordings, we can see how the proposed tracking system clearly outperforms the baseline methods that use spectrograms as inputs and that it also outperforms the 1D CNN methods when we use maps with at least 16x32 resolution.

According to [23], the reverberation time of the room where the recordings were performed was $T_{60} \approx 0.5\,\text{s}$, so we can observe some degradation in the performance of Cross3D when it is used over low resolution power maps compared with the results obtained with the simulated dataset (see Fig. 8), but it disappears when the resolution of the maps increases; actually we even reach lower errors in the LOCATA dataset than with the simulated test dataset. Using a 1D CNN also suffers a similar degradation, but its most dramatic impact is on the methods which use spectrograms as inputs. In contrast, the use of a 1D CNN over the coordinates of the maximums of high resolution SRP-PHAT maps does not suffer almost any degradation; but it may not be an interesting approach since, having computed the 64x128 resolution maps, we can obtain far better results using the whole maps as inputs of Cross3D.

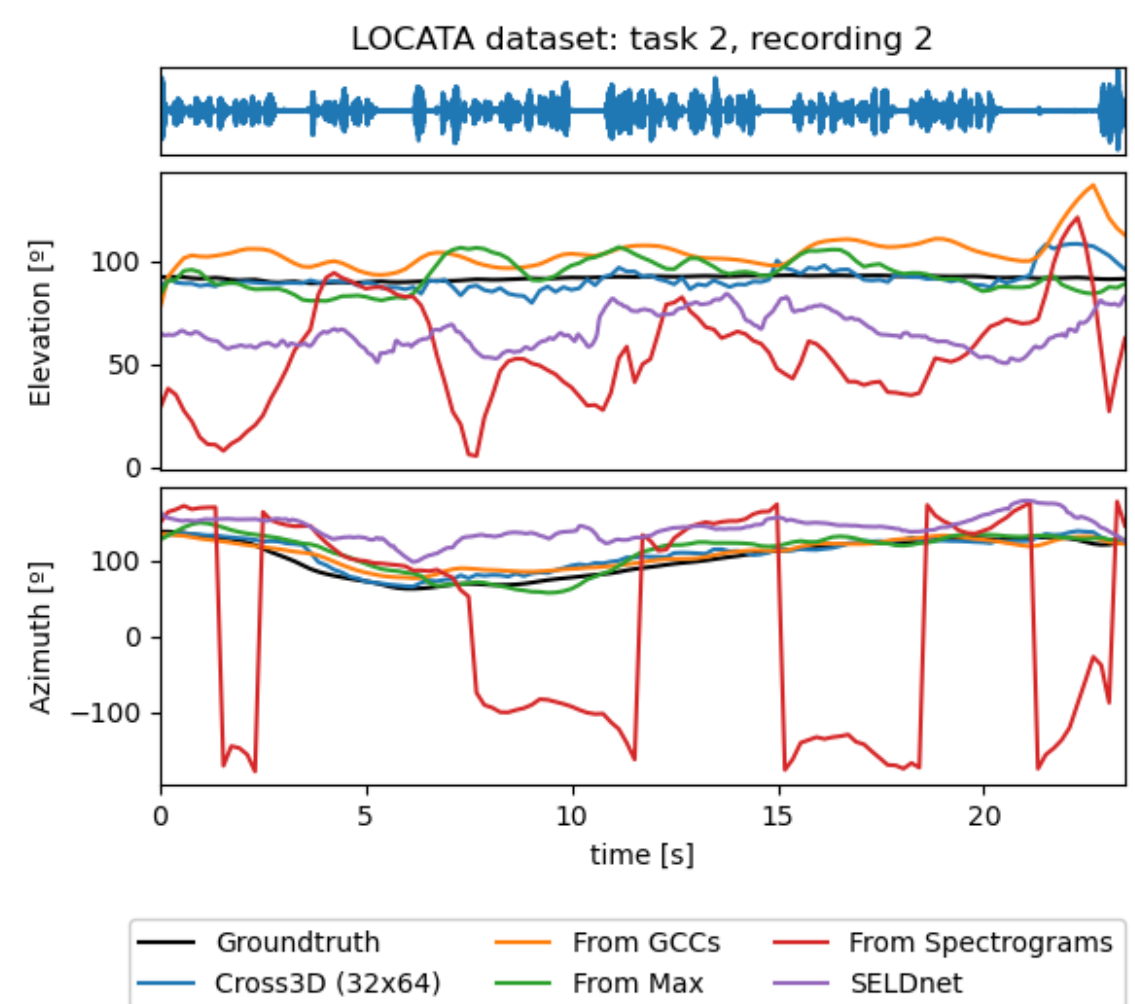

Fig. 9. DOA estimated for the second recording of the second task of the LOCATA challenge using maps with Cross3D over 32x64 maps and the baseline methods.

As an example, Fig. 9 shows the DOA estimation of the second recording of the third task of the LOCATA dataset, where all the methods obtained a RMSAE quite close to their average. We can see how Cross3D performs the best estimation of the analyzed methods both for the elevation and for the azimuth. We can also see that the LOCATA dataset has longer silences than the ones present in the simulated dataset, which could also explain why some of the methods obtained lower results with this dataset. In order to make the methods based on CNNs more robust against longer silences, we should include them in the simulation of the training dataset and, probably, increase the temporal receptive field of the models, which could be done increasing the number of layers, the temporal size of its kernels, or including longer temporal dilations in the convolutions.

## VI. Conclusions

In this paper, we have presented a new sound source DOA estimation and tracking system based on the well known SRP-PHAT method and a three-dimensional Convolutional Neural Network. The use of a fully causal convolutional architecture, without any bidirectional recurrent layer, makes our proposal feasible for real-time applications, being able to provide a new DOA estimation each $192\,\mathrm{ms}$. We used a 3D CNN over time sequences of elevation and azimuth maps computed from the signals captured by a compact array but, using distributed arrays to compute 3D maps, the extension of the technique to use a 4D CNN would be straightforward.

The experiments performed show that the SRP-PHAT maps are a good input feature to be used in tracking systems based on deep learning, being much more robust to reverberation and noise than the use of spectrograms as proposed in most of the recent literature. They also prove that it is possible to obtain a good tracking performance using only causal convolutional layers and that non-causal recurrent layers are not needed.

Due to the difficulty of recording a hand-labeled dataset of moving sources large enough to train a neural network, we have introduced a new procedure for generating random trajectories and simulate them as they are needed for training. With it, we have a infinite size dataset whose parameters can be easily modified during training to accelerate the convergence or during test to analyze the performance of the model in specific scenarios. To prove that the models trained with this procedure are general enough to deal with actual recordings, we have tested our model with the LOCATA dataset and obtained satisfactory results.

As a baseline method for our main proposal, we have also introduced a new architecture, based on the use of a causal 1D CNN over the GCCs, that also presents a good performance and robustness and that may be interesting for applications where the computation of the SRP-PHAT maps is not possible due to computational resource limitations.

## Acknowledgment

This work was supported in part by the Regional Government of Aragon (Spain) with a grant for postgraduate research contracts (2017-2021) co-founded by the Operative Program FSE Aragon 2014-2020.

This material is based upon work supported by Google Cloud.

## References

[1] C. Rascon and I. Meza, "Localization of sound sources in robotics: A review," *Robotics and Autonomous Systems*, vol. 96, pp. 184–210, 2017.
[2] V. Tourbabin and B. Rafaely, "Theoretical Framework for the Optimization of Microphone Array Configuration for Humanoid Robot Audition," *IEEE/ACM Transactions on Audio, Speech, and Language Processing*, vol. 22, no. 12, pp. 1803–1814, Dec. 2014.
[3] A. Farina and L. Tronchin, "3D Sound Characterisation in Theatres Employing Microphone Arrays," *Acta Acustica united with Acustica*, vol. 99, no. 1, pp. 118–125, Jan. 2013.
[4] K. Kumatani, J. McDonough, and B. Raj, "Microphone Array Processing for Distant Speech Recognition: From Close-Talking Microphones to Far-Field Sensors," *IEEE Signal Processing Magazine*, vol. 29, no. 6, pp. 127–140, Nov. 2012.
[5] C. Spille, B. Kollmeier, and B. T. Meyer, "Comparing human and automatic speech recognition in simple and complex acoustic scenes," *Computer Speech & Language*, vol. 52, pp. 123–140, 2018.
[6] R. Ma, G. Liu, Q. Hao, and C. Wang, "Smart microphone array design for speech enhancement in financial VR and AR," in *2017 IEEE SENSORS*, Oct. 2017, pp. 1–3.
[7] C. Knapp and G. Carter, "The generalized correlation method for estimation of time delay," *IEEE Transactions on Acoustics, Speech, and Signal Processing*, vol. 24, no. 4, pp. 320–327, Aug. 1976.
[8] J. Benesty, "Adaptive eigenvalue decomposition algorithm for passive acoustic source localization," *The Journal of the Acoustical Society of America*, vol. 107, no. 1, pp. 384–391, Dec. 1999.
[9] L. Comanducci, M. Cobos, F. Antonacci, and A. Sarti, "Time Difference of Arrival Estimation from Frequency-Sliding Generalized Cross-Correlations Using Convolutional Neural Networks," in *ICASSP 2020 - 2020 IEEE International Conference on Acoustics, Speech and Signal Processing (ICASSP)*, May 2020, pp. 4945–4949.
[10] J. H. DiBiase, "A high-accuracy, low-latency technique for talker localization in reverberant environments using microphone arrays," Ph.D. dissertation, Brown University, 2000.
[11] J. H. DiBiase, H. F. Silverman, and M. Brandstein, "Robust Localization in Reverberant Rooms," in *Microphone Arrays: Signal Processing Techniques and Applications*. Berlin, Heidelberg: Springer Berlin Heidelberg, 2001.
[12] R. Schmidt, "Multiple emitter location and signal parameter estimation," *IEEE Transactions on Antennas and Propagation*, vol. 34, no. 3, pp. 276–280, Mar. 1986.
[13] J. P. Dmochowski, J. Benesty, and S. Affes, "Broadband Music: Opportunities and Challenges for Multiple Source Localization," in *2007 IEEE Workshop on Applications of Signal Processing to Audio and Acoustics*, Oct. 2007, pp. 18–21.
[14] J. Traa and P. Smaragdis, "A Wrapped Kalman Filter for Azimuthal Speaker Tracking," *IEEE Signal Processing Letters*, vol. 20, no. 12, pp. 1257–1260, Dec. 2013.
[15] Y. Tian, Z. Chen, and F. Yin, "Distributed Kalman filter-based speaker tracking in microphone array networks," *Applied Acoustics*, vol. 89, pp. 71–77, Mar. 2015.
[16] D. B. Ward, E. A. Lehmann, and R. C. Williamson, "Particle filtering algorithms for tracking an acoustic source in a reverberant environment," *IEEE Transactions on Speech and Audio Processing*, vol. 11, no. 6, pp. 826–836, Nov. 2003.
[17] W.-K. Ma, B.-N. Vo, S. S. Singh, and A. Baddeley, "Tracking an unknown time-varying number of speakers using TDOA measurements: A random finite set approach," *IEEE Transactions on Signal Processing*, vol. 54, no. 9, pp. 3291–3304, Sep. 2006.
[18] X. Xiao, S. Zhao, X. Zhong, D. L. Jones, E. S. Chng, and H. Li, "A learning-based approach to direction of arrival estimation in noisy and reverberant environments," in *2015 IEEE International Conference on Acoustics, Speech and Signal Processing (ICASSP)*, Apr. 2015, pp. 2814–2818.
[19] R. Takeda and K. Komatani, "Discriminative multiple sound source localization based on deep neural networks using independent location model," in *2016 IEEE Spoken Language Technology Workshop (SLT)*, Dec. 2016, pp. 603–609.
[20] J. M. Vera-Diaz, D. Pizarro, and J. Macias-Guarasa, "Towards End-to-End Acoustic Localization Using Deep Learning: From Audio Signals to Source Position Coordinates," *Sensors*, vol. 18, no. 10, p. 3418, Oct. 2018.

[21] E. L. Ferguson, S. B. Williams, and C. T. Jin, "Sound Source Localization in a Multipath Environment Using Convolutional Neural Networks," in *2018 IEEE International Conference on Acoustics, Speech and Signal Processing (ICASSP)*, Apr. 2018, pp. 2386–2390.

[22] S. Adavanne, A. Politis, J. Nikunen, and T. Virtanen, "Sound Event Localization and Detection of Overlapping Sources Using Convolutional Recurrent Neural Networks," *IEEE Journal of Selected Topics in Signal Processing*, vol. 13, no. 1, pp. 34–48, Mar. 2019.

[23] H. W. Löllmann, C. Evers, A. Schmidt, H. Mellmann, H. Barfuss, P. A. Naylor, and W. Kellermann, "The LOCATA Challenge Data Corpus for Acoustic Source Localization and Tracking," in *2018 IEEE 10th Sensor Array and Multichannel Signal Processing Workshop (SAM)*, Jul. 2018, pp. 410–414.

[24] H. Do and H. F. Silverman, "A Fast Microphone Array SRP-PHAT Source Location Implementation using Coarse-To-Fine Region Contraction(CFRC)," in *2007 IEEE Workshop on Applications of Signal Processing to Audio and Acoustics*, Oct. 2007, pp. 295–298.

[25] ——, "Stochastic particle filtering: A fast SRP-PHAT single source localization algorithm," in *2009 IEEE Workshop on Applications of Signal Processing to Audio and Acoustics*, Oct. 2009, pp. 213–216.

[26] L. O. Nunes, W. A. Martins, M. V. S. Lima, L. W. P. Biscainho, M. V. M. Costa, F. M. Gonçalves, A. Said, and B. Lee, "A Steered-Response Power Algorithm Employing Hierarchical Search for Acoustic Source Localization Using Microphone Arrays," *IEEE Transactions on Signal Processing*, vol. 62, no. 19, pp. 5171–5183, Oct. 2014.

[27] M. Cobos, A. Marti, and J. J. Lopez, "A Modified SRP-PHAT Functional for Robust Real-Time Sound Source Localization With Scalable Spatial Sampling," *IEEE Signal Processing Letters*, vol. 18, no. 1, pp. 71–74, Jan. 2011.

[28] C. Evers, Y. Dorfan, S. Gannot, and P. A. Naylor, "Source tracking using moving microphone arrays for robot audition," in *2017 IEEE International Conference on Acoustics, Speech and Signal Processing (ICASSP)*, Mar. 2017, pp. 6145–6149.

[29] O. Schwartz, Y. Dorfan, E. A. P. Habets, and S. Gannot, "Multi-speaker DOA estimation in reverberation conditions using expectation-maximization," in *2016 IEEE International Workshop on Acoustic Signal Enhancement (IWAENC)*, Sep. 2016, pp. 1–5.

[30] S. Chakrabarty and E. A. P. Habets, "Multi-Speaker DOA Estimation Using Deep Convolutional Networks Trained With Noise Signals," *IEEE Journal of Selected Topics in Signal Processing*, vol. 13, no. 1, pp. 8–21, Mar. 2019.

[31] S. Adavanne, A. Politis, and T. Virtanen, "Direction of Arrival Estimation for Multiple Sound Sources Using Convolutional Recurrent Neural Network," in *2018 26th European Signal Processing Conference (EUSIPCO)*, Sep. 2018, pp. 1462–1466.

[32] N. Yalta, K. Nakadai, and T. Ogata, "Sound Source Localization Using Deep Learning Models," *Journal of Robotics and Mechatronics*, vol. 29, no. 1, pp. 37–48, 2017.

[33] Y. Sun, J. Chen, C. Yuen, and S. Rahardja, "Indoor Sound Source Localization With Probabilistic Neural Network," *IEEE Transactions on Industrial Electronics*, vol. 65, no. 8, pp. 6403–6413, Aug. 2018.

[34] W. He, P. Motlicek, and J.-M. Odobez, "Deep Neural Networks for Multiple Speaker Detection and Localization," in *2018 IEEE International Conference on Robotics and Automation (ICRA)*, May 2018, pp. 74–79.

[35] S. Kapka and M. Lewandowski, "Sound Source Detection, Localization and Classification using Consecutive Ensemble of CRNN Models," in *Proceedings of the Detection and Classification of Acoustic Scenes and Events 2019 Workshop (DCASE2019)*. New York University, 2019, pp. 119–123.

[36] H. Cordourier, P. Lopez Meyer, J. Huang, J. Del Hoyo Ontiveros, and H. Lu, "GCC-PHAT Cross-Correlation Audio Features for Simultaneous Sound Event Localization and Detection (SELD) on Multiple Rooms," in *Proceedings of the Detection and Classification of Acoustic Scenes and Events 2019 Workshop (DCASE2019)*. New York University, Oct. 2019.

[37] Y. Cao, T. Iqbal, Q. Kong, M. B. Galindo, W. Wang, and M. D. Plumbley, "TWO-STAGE SOUND EVENT LOCALIZATION AND DETECTION USING INTENSITY VECTOR AND GENERALIZED CROSS-CORRELATION," in *Proceedings of the Detection and Classification of Acoustic Scenes and Events 2019 Workshop (DCASE2019)*. New York University, 2019.

[38] L. Perotin, A. Défossez, E. Vincent, R. Serizel, and A. Guérin, "Regression versus classification for neural network based audio source localization," *IEEE Workshop on Applications of Signal Processing to Audio and Acoustics*, p. 6, 2019.

[39] Z. Tang, J. D. Kanu, K. Hogan, and D. Manocha, "Regression and Classification for Direction-of-Arrival Estimation with Convolutional Recurrent Neural Networks," in *Interspeech 2019*. ISCA, Sep. 2019, pp. 654–658.

[40] D. Diaz-Guerra and J. R. Beltran, "Source cancellation in cross-correlation functions for broadband multisource DOA estimation," *Signal Processing*, vol. 170, p. 107442, May 2020.

[41] A. Brutti, M. Omologo, and P. Svaizer, "Multiple Source Localization Based on Acoustic Map De-Emphasis," *EURASIP Journal on Audio, Speech, and Music Processing*, vol. 2010, no. 1, p. 147495, Dec. 2010.

[42] D. Diaz-Guerra and J. R. Beltran, "Direction of Arrival Estimation with Microphone Arrays Using SRP-PHAT and Neural Networks," in *2018 IEEE 10th Sensor Array and Multichannel Signal Processing Workshop (SAM)*, Jul. 2018, pp. 617–621.

[43] L. Perotin, R. Serizel, E. Vincent, and A. Guérin, "CRNN-Based Multiple DoA Estimation Using Acoustic Intensity Features for Ambisonics Recordings," *IEEE Journal of Selected Topics in Signal Processing*, vol. 13, no. 1, pp. 22–33, Mar. 2019.

[44] T. S. Cohen, M. Geiger, J. Köhler, and M. Welling, "Spherical CNNs," in *International Conference on Learning Representations*, 2018.

[45] Y. Bengio, J. Louradour, R. Collobert, and J. Weston, "Curriculum learning," in *Proceedings of the 26th Annual International Conference on Machine Learning*, ser. ICML '09. Montreal, Quebec, Canada: Association for Computing Machinery, Jun. 2009, pp. 41–48.

[46] V. Panayotov, G. Chen, D. Povey, and S. Khudanpur, "Librispeech: An ASR corpus based on public domain audio books," in *2015 IEEE International Conference on Acoustics, Speech and Signal Processing (ICASSP)*, Apr. 2015, pp. 5206–5210.

[47] J. Wiseman, "Wiseman/py-webrtcvad," Nov. 2019.

[48] J. B. Allen and D. A. Berkley, "Image method for efficiently simulating small-room acoustics," *The Journal of the Acoustical Society of America*, vol. 65, no. 4, pp. 943–950, Apr. 1979.

[49] D. Diaz-Guerra, A. Miguel, and J. R. Beltran, "gpuRIR: A python library for room impulse response simulation with GPU acceleration," *Multimedia Tools and Applications*, Oct. 2020.

[50] H. W. Löllmann, C. Evers, A. Schmidt, H. Mellmann, H. Barfuss, P. A. Naylor, and W. Kellermann, "IEEE-AASP Challenge on Acoustic Source Localization and Tracking: Documentation of Final Release," https://locata.lms.tf.fau.de/datasets/, Jan. 2020.

[51] K. He, X. Zhang, S. Ren, and J. Sun, "Delving Deep into Rectifiers: Surpassing Human-Level Performance on ImageNet Classification," in *2015 IEEE International Conference on Computer Vision (ICCV)*, Dec. 2015, pp. 1026–1034.

[52] D. P. Kingma and J. Ba, "Adam: A Method for Stochastic Optimization," in *ICLR 2015*, San Diego.

[53] A. Paszke, S. Gross, F. Massa, A. Lerer, J. Bradbury, G. Chanan, T. Killeen, Z. Lin, N. Gimelshein, L. Antiga, A. Desmaison, A. Kopf, E. Yang, Z. DeVito, M. Raison, A. Tejani, S. Chilamkurthy, B. Steiner, L. Fang, J. Bai, and S. Chintala, "PyTorch: An Imperative Style, High-Performance Deep Learning Library," in *Advances in Neural Information Processing Systems 32*, H. Wallach, H. Larochelle, A. Beygelzimer, F. d'Alché-Buc, E. Fox, and R. Garnett, Eds. Curran Associates, Inc., 2019, pp. 8026–8037.

[54] K. Cho, B. van Merriënboer, C. Gulcehre, D. Bahdanau, F. Bougares, H. Schwenk, and Y. Bengio, "Learning Phrase Representations using RNN Encoder–Decoder for Statistical Machine Translation," in *Proceedings of the 2014 Conference on Empirical Methods in Natural Language Processing (EMNLP)*. Doha, Qatar: Association for Computational Linguistics, Oct. 2014, pp. 1724–1734.